\documentclass[aps,prb,twocolumn,showpacs, 10pt]{revtex4-1}
\usepackage{graphicx}
\usepackage{amsmath}
\usepackage{amssymb}
\usepackage{color}

\newcommand{\Eu}{Eu(Fe$_{0.93}$Rh$_{0.07}$)$_2$As$_2$}

\begin{document} 
\title{Effects of pressure and magnetic field on the re-entrant superconductor \Eu}
\author{A. L\"ohle}
\affiliation{1.~Physikalisches Institut, Universit\"{a}t
Stuttgart, Pfaffenwaldring 57, D-70569 Stuttgart Germany}
\author{A. Baumgartner}
\affiliation{1.~Physikalisches Institut, Universit\"{a}t
Stuttgart, Pfaffenwaldring 57, D-70569 Stuttgart Germany}
\author{S. Zapf}
\affiliation{1.~Physikalisches Institut, Universit\"{a}t
Stuttgart, Pfaffenwaldring 57, D-70569 Stuttgart Germany}
\author{W. H. Jiao}
\affiliation{Zhejiang University, Hangzhou, Zhejiang, China}
\author{G. H. Cao}
\affiliation{Zhejiang University, Hangzhou, Zhejiang, China}
\author{M. Dressel}
\affiliation{1.~Physikalisches Institut, Universit\"{a}t
Stuttgart, Pfaffenwaldring 57, D-70569 Stuttgart Germany}
\date{\today}
\begin{abstract}
Electron-doped \Eu \ has been systematically studied by high pressure investigations of the magnetic and electrical transport properties, in order to unravel the complex interplay of superconductivity and magnetism. The compound reveals an exceedingly broad re-entrant transition to the superconducting state between $T_{\rm{c,on}} = 19.8$~K and $T_{\rm{c,0}} = 5.2$~K due to a canted A-type antiferromagnetic ordering of the Eu$^{2+}$ moments at $T_{\rm{N}} = 16.6$~K and a re-entrant spin glass transition at $T_{\rm{SG}} = 14.1$~K. At ambient pressure evidences for the coexistence of superconductivity and ferromagnetism could be observed, as well as a magnetic-field-induced enhancement of the zero-resistance temperature $T_{\rm{c,0}}$ up to $7.2$~K with small magnetic fields applied parallel to the \textit{ab}-plane of the crystal. We attribute the field-induced-enhancement of superconductivity to the suppression of the ferromagnetic component of the Eu$^{2+}$ moments along the \textit{c}-axis, which leads to a reduction of the orbital pair breaking effect. Application of hydrostatic pressure suppresses the superconducting state around $14$~kbar along with a linear temperature dependence of the resistivity, implying that a non-Fermi liquid region is located at the boundary of the superconducting phase. At intermediate pressure, an additional feature in the resistivity curves is identified, which can be suppressed by external magnetic fields and competes with the superconducting phase. 
We suggest that the effect of negative pressure by the chemical Rh substitution in \Eu \ is partially reversed, leading to a re-activation of the spin density wave.

\end{abstract}


\maketitle
%
%
\section{Introduction}
\label{introduction}
The discovery of high-$T_c$ superconductivity in the iron-pnictide family \cite{Kamihara06} has been one of the most exciting recent developments  in condensed matter research. While magnetism and superconductivity are traditionally competitive phenomena in conventional superconductors \cite{Ginzburg56, Saint69}, there is growing experimental evidence that the unconventional superconductivity of the iron-pnictides is closely linked to magnetism as also supposed for the high-$T_c$ cuprates, which are still the most intensively investigated unconventional superconductors \cite{Moriya90, Monthoux92}. Among the parent compounds of iron-based superconductors the Eu containing 122-compound \cite{Jeevan08, Ren08} stands out in particular due to its magnetic sub-lattice formed by the Eu$^{2+}$ ions, carrying a local moment of $S = 7/2$ \cite{Jiang09, Xiao09, Herrero09, Zapf16}. In this case -– additional to the spin-density-wave (SDW) ordering in the FeAs layers at approximately $190$~K -– the Eu$^{2+}$ moments align ferromagnetically along the $a$-axis and antiferromagnetically along the $c$-axis in a so called A-type antiferromagnetic pattern \cite{Zapf13} at $T_{\rm{N}} \approx 19$~K. Application of hydrostatic or chemical pressure \cite{Terashima09, Jeevan11} as well as hole \cite{Jeevan082} and electron \cite{Jiang092} doping of EuFe$_2$As$_2$ lead to a suppression of the SDW state and an emergence of superconductivity with transition temperatures up to $40$~K. In those compounds, where the energy scales of the magnetic and superconducting state are in close proximity, the competition between the two orders become very apparent as exotic effects such as a resistivity re-entrance around $T_{\rm{N}}$ \cite{He10} or an enhancement of $T_c$ by the application of small external magnetic fields \cite{Tran12}. Up to now considerable effort was put into developing a clear picture of how superconductivity can coexist or even be induced by magnetic effects. In case of the strong Eu$^{2+}$ magnetism an additional re-entrant spin glass (SG) phase \cite{Zapf13} seems to be a key in understanding the coexistence, whereas in general the idea of a magnetic quantum critical point lying beneath the superconducting dome has been a long-standing hypothesis \cite{Shibauchi14}. 
In this study, detailed magneto-transport measurements under hydrostatic pressure up to $18$~kbar together with systematic dc and ac magnetization measurements were employed in order to investigate this interesting entanglement of superconductivity and magnetism in the phase diagram of an nearly optimal doped member of the rare earth EuFe$_2$As$_2$ iron pnictide family.


\section{Experiment}
\label{experiment}
\begin{figure}
\centering
\includegraphics
[width=0.9\columnwidth]
{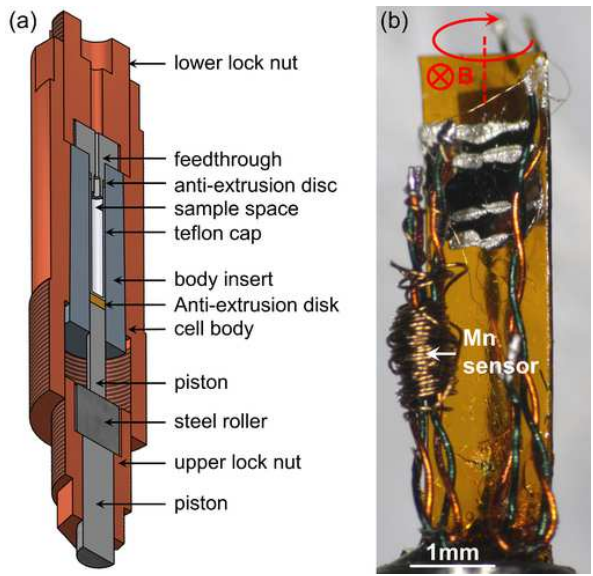}
\caption{(Color online) (a) Technical drawing of the clamp piston pressure cell utilized for this study. The feedthrough is equipped with 6 pairs of twisted wire pairs for the electrical measurements. (b) Microscope picture of the \Eu\ sample mounted on the kapton sample stage of the pressure cell feedthrough. The magnetic field direction as well as the rotation axis for the angle dependent magneto tansport measurements are marked in red, while the Mn wire coil for the in-situ pressure determination is indicated in white.}
\label{figure1}
\end{figure}

Single crystals of Eu(Fe$_{1-x}$Rh$_x$)$_2$As$_2$ were grown via the self-flux method \cite{Neubauer16, Cao09, Cao13}. The quality and precise chemical composition of the crystals was subsequently checked by energy-dispersive x-ray spectroscopy (EDX) on several points of the crystal. The crystal used for the pressure dependent investigations [see Fig. \ref{figure1} (b)] showed no evidences of impurity elements in the EDX spectra and the chemical composition corresponds to a $x=7$~\% rhodium doping on the iron side. For determining the transport and magnetic properties of the Rh -- i.e. electron doped -- sample, detailed temperature and magnetic field dependent four-point dc-resistivity measurements were carried out. The resistivity was thereby determined using electric currents up to $5$~mA -– which ensures a linear response -– from room temperature down to $1.8$~K with a relative low cooling rate of $0.3$~K/min to guarantee a proper thermalization of the sample. At ambient pressure additional ac and dc magnetization measurement were performed  on a Quantum Design Magnetic Property Measurement System (MPMS) to unambiguously identify the origin of the distinct features of the dc-resistivity curve and to get further information about the underlying physics.

The pressure dependent electrical resistivity measurements were carried out by using a clamp piston pressure cell sketched in Figure \ref{figure1} with Daphne oil 7373 \cite{Yokogawa07} as the pressure transmitting medium. For the in-plane resistivity measurements, four stripe-like contacts were made by silver paint on the sample, connecting it to the twisted wire pairs of the pressure cell feedthrough with 10~$\mu$m thick gold wires. To provide the opportunity of magnetoresistance measurements with applied in- and out-of-plane magnetic fields up to 65~kG, the sample was mounted on the kapton sample stage of the electrical feedthrough as shown in Figure \ref{figure1} (b). In this configuration the sample can be rotated together with the whole pressure cell with respect to the horizontal magnetic field [as indicated by the red arrow in Fig. \ref{figure1} (b)]. For determining the actual pressure inside of the cell, a Mn pressure sensor in form of a wire coil as shown in Figure \ref{figure1} (b) was used. In this way the pressure could be measured in-situ and therefore the pressure loss during the cooling process could be taken into account in the further analysis of the data. 


\section{Results}
\label{results}

\subsection{Electronic and magnetic properties at ambient conditions}
\begin{figure}
\centering
\includegraphics
[width=0.9\columnwidth]
{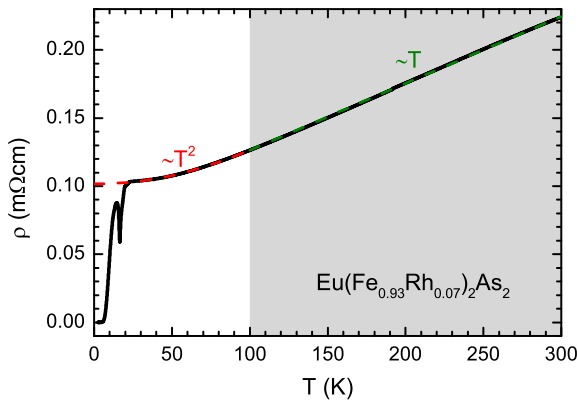}
\caption{(Color online) Temperature dependent dc resistivity data of \Eu\ at ambient pressure. In the high-temperature regime (grey shaded area) above $100$~K, the resistivity shows a linear behavior in $T$ (green dashed line), while the low temperature region, above the superconducting transition, can be fitted by the power law term $\rho(T)=\rho_0+c\cdot T^n$ (red dashed curve) with an exponent $n = 2$.}
\label{figure2}
\end{figure}

\begin{figure}[ht]
\centering
\includegraphics
[width=0.9\columnwidth]
{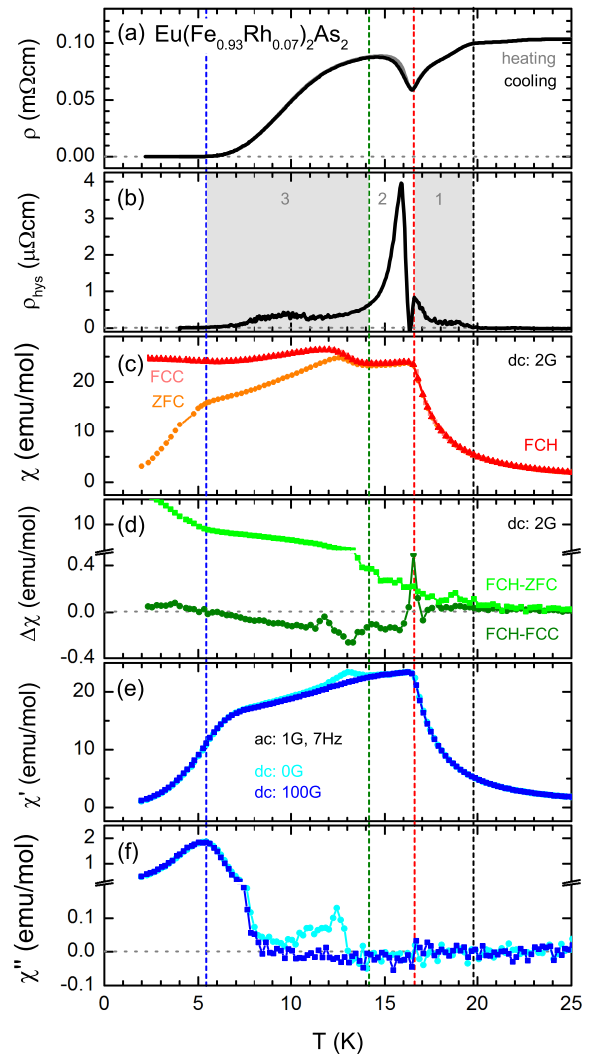}
\caption{(Color online) Comparison between the low temperature dc resistivity and magnetization data of \Eu \ recorded at ambient pressure. In panel (a) the dc-resistivity with its clear re-entrant superconducting behavior is shown, while panel (b) displays the thermal hysteresis of the resistivity occurring during the cooling-heating-cycle. Panels (c)-(f) illustrate the dc and ac magnetic susceptibility. The dashed lines are guides to the eyes and mark the different transitions namely the onset of superconductivity ($T_{\rm{c,on}}= 19.8$~K, dashed black line), the antiferromagnetic ordering of the Eu$^{2+}$ moments ($T_{\rm{N}}= 16,6$~K, dashed red line), the re-entrant spin glass ($T_{\rm{SG}}= 14,1$~K, dashed green line) and the approach of zero resistance ($T_{\rm{c,0}}= 5.4$~K, dashed blue line).}
\label{figure3}
\end{figure}

Figure \ref{figure2} shows the temperature dependence of the resistivity measured along the \textit{ab}-plane from room temperature down to $2$~K. In the high temperature region (grey shaded area) above $100$~K, the resistivity shows a $T$-linear behavior as expected for a normal metal with dominant electron-phonon scattering. At low temperatures, but still above the superconducting transition, the resistivity can be fitted with the power law expression 

\begin{equation}
\rho(T) = \rho_0 + a \cdot T^n
\label{eq:rho}
\end{equation}

leading to an exponent $n=2$ which points out Fermi-liquid behavior with appreciable electron-electron interaction. Due to the missing indication of any SDW associated anomaly in the resistivity curve, the onset of superconductivity around $19.8$~K (see Fig. \ref{figure3}), together with the fact that for optimal doping a $T$-linear behavior down to the lowest temperature is expected \cite{Tam13, Kurita13} we suggest the \Eu \ sample to be located in the slightly under-doped region of the phase diagram (cf. $x=9$~\% rhodium doping in \cite{Jiao16}). The residual resistivity ratio of

\begin{equation}
RRR=\frac{\rho_{300K}}{\rho_{0K,fit}} = 2.2
\label{eq:RRR}
\end{equation}

indicates thereby a good quality of the sample. 

Panel (a) in Figure \ref{figure3} depicts the low-temperature resistivity curves of \Eu \ -- measured in a cooling-heating cycle -- which show a clear re-entrant superconducting transition as well as a small thermal hysteresis exhibiting a distinct three-stage shape [see Fig. \ref{figure3} panel (b)]. 

To identify the underlying physical processes of the distinct features of the resistivity curve, dc and ac magnetization measurements (panels (c)-(f) in Fig. \ref{figure3}) are utilized based on Ref. \onlinecite{Zapf13, Baumgartner16}. The first kink in the resistivity curve coincides with the opening of the hysteresis at $T_{\rm{c,on}} = 19.6$~K; it is caused by the onset of superconductivity, as can be seen by the splitting of the zero field cooled (ZFC) and field cooled heat (FCH) susceptibility curves [Fig. \ref{figure3} (c) and (d), black dashed line]. The local minimum of $\rho(T)$ together with the onset of the second peak of the hysteresis curve at $T_{\rm{N}} = 16.6$~K occur simultaneously with a clear kink in the dc and ac magnetization data (Fig. \ref{figure3} panel (c) and (e), red dashed line) which corresponds to a canted A-type anti-ferromagnetic ordering of the local Eu$^{2+}$ moments \cite{Jiang092, Zapf11}. Lowering the temperature further leads to a second magnetic transition at $T_{\rm{SG}} = 14.1$~K into a re-entrant spin glass phase, as identified  by Zapf \textit{et al.} \cite{Zapf13, Zapf11} in P substituted samples. The typical features of this phase can be preferably seen in the ac magnetization data $\chi'_{ac}$($T$) and $\chi''_{ac}$($T$) (panel (e) and (f) in Fig. \ref{figure3}, green dashed line) and coincide with the local maximum of the re-entrant superconducting transition in the resistivity curve. At $T_{\rm{c,0}} = 5.4$~K, the superconducting transition is completed and the sample reaches zero resistance (blue dashed line in Fig. \ref{figure3}), while the hysteresis closes and the magnetization shows a distinct kink in the ZFC dc susceptibility and a maximum in $\chi''$($T$). \\ 
In the following analysis we investigate the effect of an external magnetic field and hydrostatic pressure on the features we identified in the \Eu \ sample.

\subsection{Magnetic field dependences of the low temperature phases}
By applying an external magnetic field we can affect the magnetic order and superconducting transition temperature of \Eu \ and thus vary the interplay between this phases. Application of in- ($H\parallel ab$) and out-of-plane ($H\parallel c$) magnetic fields give here the possibility to selectively modify the different aspects of the magnetic ordering and therefore identify their basic role for the entire system.

\subsubsection{Magnetic ordering}
\begin{figure}
\centering
\includegraphics
[width=0.9\columnwidth]
{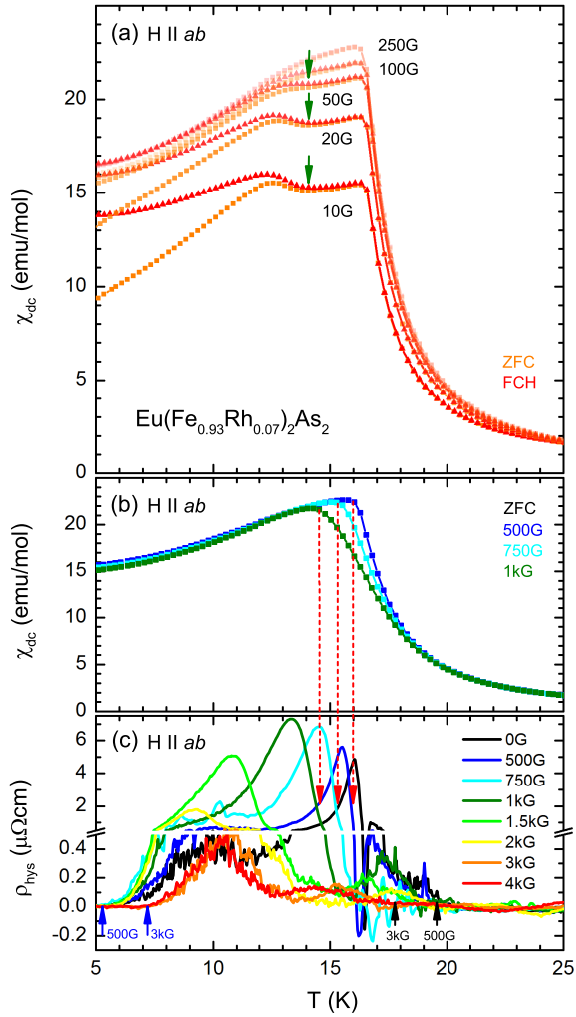}
\caption{(Color online) (a) In-plane dc magnetization data indicate that the spin glass state of \Eu \ (indicated by the green arrows) vanishes already in the presence of minor magnetic fields of around $250$~G. Comparison of the magnetization (b) data with the evolution of the thermal hysteresis in $\rho$(T) (c) for different magnetic fields applied parallel to the \textit{ab}-plane of the \Eu \ single crystal. The red dashed arrows are guides to the eye which indicate the antiferromagnetic transition temperatures. The opening of the big hysteresis feature (labeled with 2 in  Fig.F \ref{figure3}) turns out to be connected with the antiferromagnetic ordering of the Eu$^{2+}$ moments at $T_{\rm{N}}$ and can therefore identify a fast suppression of the antiferromagnetic state. The black and blue arrows indicate - exemplary for two different magnetic fields - $T_{\rm{c,on}}$ and $T_{\rm{c,0}}$ as the opening and closing of the overall hysteresis.}
\label{figure4}
\end{figure}

Figure \ref{figure4} shows low field dc susceptibility ($\chi_{dc}=M(T)/\mu_0H$) and resistivity data recorded by applied magnetic fields parallel to the \textit{ab}-plane of the \Eu \ single crystal.

\begin{figure}
\centering
\includegraphics
[width=0.9\columnwidth]
{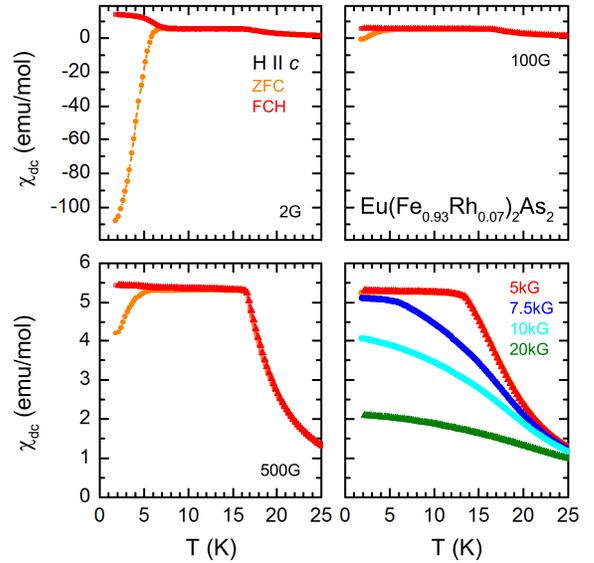}
\caption{(Color online) Temperature dependent out-of-plane ($H\parallel c$) magnetization of \Eu \ measured in ZFC-FCC-FCH cycles from $2$~K to $25$~K at applied magnetic fields up to $20$~kG. The missing kink from the magnetic ordering transition and the absence of a splitting in the ZFC-FCH data reveals a suppression of $T_{\rm{N}}$ between $7.5$~kG and $10$~kG down to the lowest measured temperature of $2$~K.}
\label{figure5}
\end{figure}

In panel (a) one can see that the feature in the magnetization data connected to the SG transition (indicated by green arrows) vanishes already in the presence of minor magnetic fields of $250$~G, while it stays at the same temperature. As the glassy freezing of the in-plane spin components in \Eu \ results from a competition of antiferromagnetic Rudeman-Kittel-Kasuya-Yosida (RKKY) interlayer coupling and the ferromagnetic intralayer spin-exchange interactions \cite{Jeevan11}, a fast suppression of this state in magnetic fields is expected. The observed behavior therefore confirms - beside the ac susceptibility measurements - the presence of a SG phase in electron-doped members of the EuFe$_2$As$_2$ family. By comparing the resistivity hysteresis with the measured susceptibility [Fig. \ref{figure4} (b) and (c)] as introduced before in the zero-field analysis, the hysteresis proves itself as a helpful tool to locate and follow the phase transitions under applied magnetic fields. The red dashed arrows indicate the antiferromagnetic transition temperatures determined by the sharp kink in the magnetization curve. The opening of the big hysteresis feature (labeled with 2 in  Fig. \ref{figure3}) confirms itself to be connected with the antiferromagnetic ordering of the Eu$^{2+}$ moments at $T_{\rm{N}}$. In line with this, one can see in Figure \ref{figure4} (c), that in-plane magnetic fields lead to a fast suppression of the antiferromagnetic ordering of the Eu$^{2+}$ moments till it vanishes completely at fields around $3$~kG. At higher fields, only a third weaker feature of the hysteresis - which is most likely connected to some vortex dynamics in the superconducting state - survives [red and orange lines in Fig. \ref{figure4} (c)]. The black and blue arrows indicate - exemplary for two different magnetic fields - $T_{\rm{c,on}}$ and $T_{\rm{c,0}}$ as the opening and closing of the overall hysteresis. It is particularly striking that the magnetic field enhances the superconducting transition temperature $T_{\rm{c,0}}$, which is untypical as superconductivity and magnetism are in general antagonistic phenomena. The origin for this anomaly and the high field evolution of the superconducting transition temperatures will be discussed in detail in the next subsection.
Figure \ref{figure5} shows the temperature dependent dc susceptibility $\chi_{dc}$, measured perpendicular to the \textit{ab}-plane ($H\parallel c$). In combination with the in-plane measurements depicted in Figure \ref{figure4}, the characteristic shape of the A-type antiferromagnetism becomes visible. While for the in-plane measurements the above discussed bump at $T_{\rm{N}} \approx 16.6$~K indicates the magnetic transition, the out-of-plane data exhibit a rather flat magnetization below $T_{\rm{N}}$. With increasing out-of-plane field, the suppression of the magnetic transition is more gradual compared to the in-plane measurements, and for fields above $7.5$~kG the kink at $T_{\rm{N}}$ gives way to a broad shoulder typical for field-induced ferromagnetism. 

\subsubsection{Superconductivity} 
\begin{figure}
\centering
\includegraphics
[width=0.9\columnwidth]
{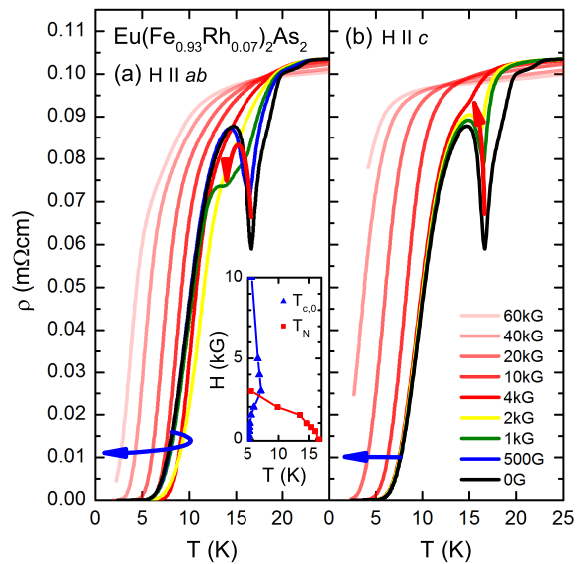}
\caption{(Color online) Magnetic field dependence of the low-temperature phases of \Eu. In-plane resistivity data from $T=2$~K to $30$~K; when measured with applied magnetic field in- and out-of plane show a clear anisotropic behavior. The in-plane magnetic field (a) leads to a suppression of the magnetic phases (red arrow) and additionally to an enhanced superconductivity at low fields (blue arrow, inset). An applied out-of-plane field (b) directly suppresses the superconductivity (blue arrow) while the magnetic transition temperatures at low fields stays nearly constant below $4$~kG (red arrow).}
\label{figure6}
\end{figure}

The layered superconducting compound \Eu \ shows a strong anisotropy of the magnetoresistance, as illustrated in Figure \ref{figure6}. In-plane magnetic fields [panel (a)] lead to a shift of the magnetic transition temperature $T_{\rm{N}}$ to smaller temperatures - as indicated by the red arrow - and finally to a complete suppression of the antiferromagnetic ordering at fields around $4$~kG. In the same field range, an enhancement of the zero-temperature point $T_{\rm{c,0}}$ from $5.2$~K up to $7.2$~K (indicated by the blue arrow) can be observed. In contrast, the out-of-plane measurements [panel (b)] just show a reduction of the antiferromagnetic impact, as can be understood by an enhanced canting of the Eu$^{2+}$ moments out of the \textit{ab}-plane. In this case no enhancement of $T_{\rm{c,0}}$ takes place.
At higher fields (around $10$~kG) finally a suppression of the superconducting state can be observed for both field directions, as expected due to spin and orbital pair-breaking \cite{Khim11, Klemm75}.

The measurements demonstrate convincingly the close interplay between the magnetic and superconducting phases. The most likely explanation for the field enhanced superconductivity with in-plane magnetic fields is a reduction of the orbital-pair-breaking effect due to a suppression of the ferromagnetic component along the \textit{c}-axis as suggested in Ref.~\onlinecite{Tran12} for Co doped EuFe$_2$As$_2$. $T_{\rm{c,on}}$ is thereby not influenced, as the magnetic transition $T_{\rm{N}}$ is at lower temperatures than the onset of superconductivity. \\

\begin{figure}
\centering
\includegraphics
[width=0.9\columnwidth]
{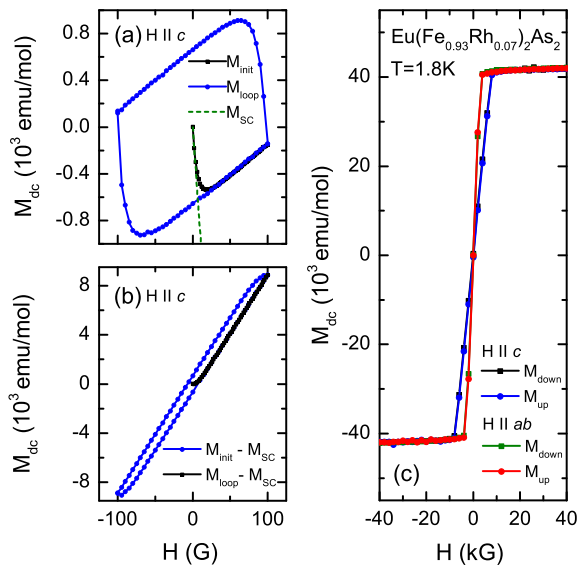}
\caption{(Color online) Isothermal magnetization measurements on \Eu \ at $T=1.8$~K for in- and out-of-plane applied external magnetic fields. (a) Hysteresis loop of the magnetization for $H\parallel c$ were $M_{\rm{init}}$ was measured after zero-field cooling. The green dashed line indicates the ideal initial magnetization $M_{\rm{SC}}$ in absence of internal magnetic fields. (b) Magnetization hysteresis loop after subtraction of $M_{\rm{SC}}$, indicating a ferromagnetic behavior of \Eu \ inside the superconducting phase. (c) Isothermal magnetization at high fields point out a saturation field of ca. $10$~kG for $H \parallel c$ and $4$~kG for $H\parallel ab$.}
\label{figure7}
\end{figure}

To address the question whether ferromagnetism along the \textit{c}-axis, due to some finite spin canting, coexists with superconductivity, isothermal magnetization measurements at $1.8$~K, well below $T_{\rm{N}}$ and $T_{\rm{c,0}}$, were employed. Figure \ref{figure7} (a) displays the hysteresis loop for $H\parallel c$. On the first glance the initial magnetization $M_{\rm{init}}$ measured after ZFC shows no linear behavior, meaning that no lower critical field $H_{\rm{c1}}^*$ can be defined; nevertheless the magnetization due to superconducting shielding effects $M_{\rm{SC}} = \xi H/(4 \pi)$ -- where $\xi$ is related to the demagnetization effect $\xi \propto 1/(1-N_d)$ -- in absence of internal magnetic fields can still be defined by the incipient slope of $M_{\rm{init}}$ \cite{Jiao13}. Subtracting $M_{\rm{SC}}$ from the initial and loop magnetization reveals a clear additional ferromagnetic hysteresis, as shown in Figure \ref{figure7} (b), indicating the coexistence of superconductivity and ferromagnetism in \Eu. The broad range magnetization at $1.8$~K is shown in Figure \ref{figure7} (c). Here one can find saturation fields of $8$~kG and $4$~kG for $H\parallel c$ and $H \parallel ab$, respectively. The saturated magnetization $M_{\rm{sat}} = 7.5$~$\mu_B$ per formula unit is consistent with a high-field induced ferromagnetic ordering of the Eu$^{2+}$ spins with a theoretical ordered moment of $gS = 7.0$~$\mu_B$ per formula unit \cite{Jin15}. Moreover, one does not observe any clear magnetization jumps at the maximum (or minimum) field, excluding that pinning effects in a hard superconductor are responsible for the hysteresis loop observed at low fields.

\subsubsection{Magnetic-field-dependent phase diagrams of \Eu}
\begin{figure}
\centering
\includegraphics
[width=0.9\columnwidth]
{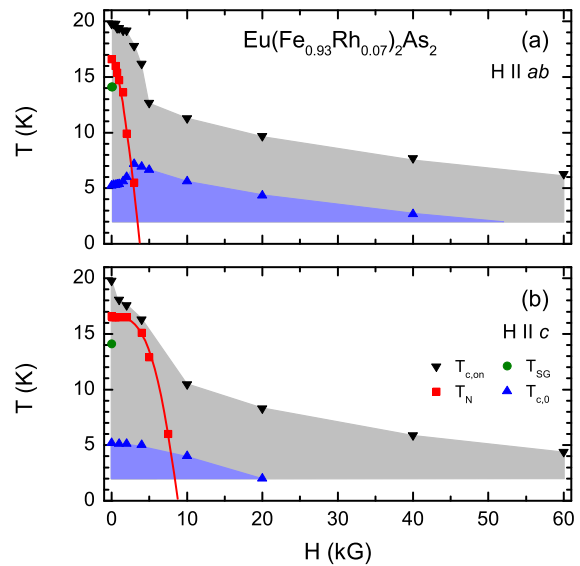}
\caption{(Color online) Magnetic field dependent phase diagrams of \Eu \ generated by the resistivity and magnetization measurements for the external magnetic field (a) $H\parallel ab$ and $H\parallel c$. The anisotropy becomes clearly visible especially in the low field region, where in-plane fields (upper panel) lead to an enhancement of $T_{\rm{c,0}}$ in combination with a fast suppression of the magnetic ordering phenomena of the Eu$^{2+}$ moments.} 
\label{figure8}
\end{figure}

 Figure \ref{figure8} shows the phase diagrams for magnetic fields parallel and perpendicular to the \textit{c}-axis of the \Eu \ crystal generated from the  resistivity and magnetization data. In these, one can identify at a glance the different aspects described above and get an insight into their interconnections. 
One can divide the phase diagrams in two main parts, at low-fields, $H < 10$~kG, the magnetic ordering phenomena greatly influence the behavior of the system; at high-fields $H > 10$~kG, the magnetic ordering is suppressed and the magnetic moments of the Eu$^{2+}$ ions are fully aligned along the field direction. At low fields, one can see the strong anisotropic behavior of the system under applied in- and out-of-plane magnetic fields, reflecting the characteristic behavior of the canted A-type antiferromagnetic ordering, even leading to a field enhanced superconductivity for $H\parallel ab$. As pointed out before, the enhancement of $T_{\rm{c,0}}$ can be explained by a reduction of the ferromagnetic component along the \textsl{c}-axis and therefore the reduction of the orbital-pair-breaking effect. The phase diagram further supports this assumption, as the enhancement takes only place until $T_{\rm{N}}$ (red dashed line) gets completely suppressed. In the high-field region, the superconducting transition shifts downwards upon increasing magnetic fields as expected from the Ginzburg-Landau theory. The suppression is thereby slightly faster for $H\parallel c$ as known for layered systems. One can estimate by a rough linear extrapolation upper critical fields of $H_{c2}^{ab}\approx 120$~kG and $H_{c2}^c\approx 100$~kG. Compared to compounds with similar $T_c$ values containing non-magnetic Ba instead Eu, these values are strikingly lower \cite{Yamamoto09}, which can be explained by a strong internal exchange field of the Eu sublattice of up to $300$~kG \cite{Nowik11}.

\subsection{Pressure-dependent transport studies on \Eu}
\begin{figure*}
\centering
\includegraphics
[width=1.8\columnwidth]
{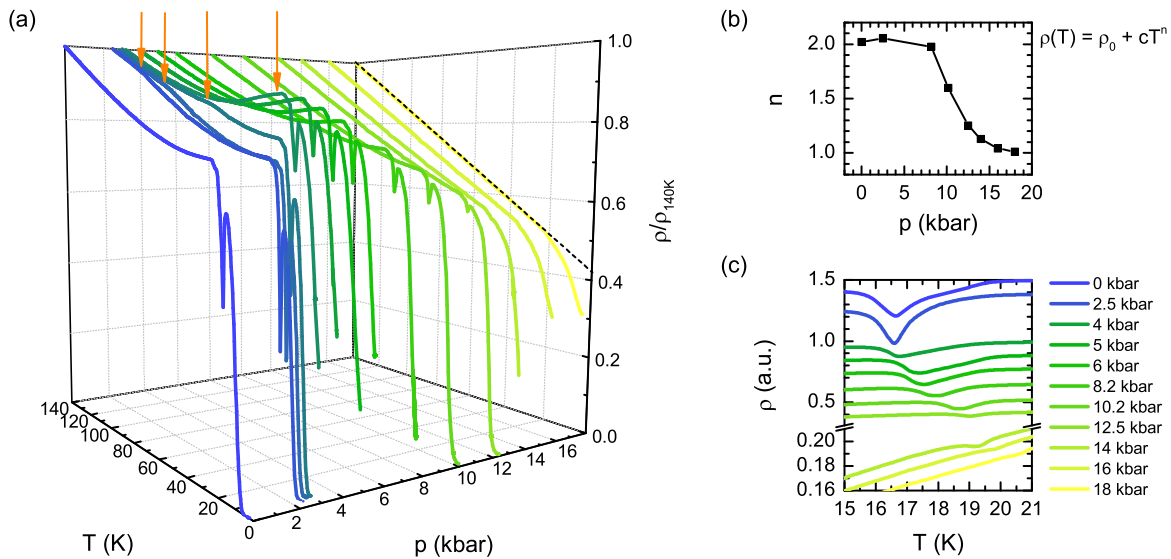}
\caption{(Color online) Pressure dependent resistivity measurements of \Eu \ up to $18$~kbar. (a) shows the resistivity data at all measured pressures in a temperature range from $2$~K up to $140$~K, normalized to their $140$~K value. An additional feature is appearing in the $\rho(T)$ curve (black arrows) at low pressures, leading to an enhancement of $T_{\rm{c,0}}$ in an intermediate pressure range. At pressures above $16$~kbar, the superconducting transition gets completely suppressed and a linear $T$ dependency can be seen (black dashed line). Panel (b) depicts the result of the power law fitting $\rho (T) = \rho_0 + cT^n$, indicating non-Fermi-liquid like behavior at the edge of the superconducting dome. (c) Zoom-in to the temperature range of the Eu$^{2+}$ ordering, revealing a clear increase of the antiferromagnetic transition temperature with increasing pressure. (For clarity the curves are shifted with respect to each other.)} 
\label{figure9}
\end{figure*}

Figure \ref{figure9} (a) shows an overview of all accomplished in-plane resistivity measurements on \Eu \ up to pressures of $18$~kbar. In the three-dimensional illustration, where $\rho(T)$ is plotted over the room temperature pressure value, one can see distinct changes of the behavior at ambient as well as at low temperatures; the exact progression will be discussed in the following paragraphs.
  
\subsubsection{Low temperature phases}
As there is a strong interplay between the low-temperature magnetic ordering of the Eu$^{2+}$ moments and the appearance of superconductivity in \Eu, let us look at the peculiarities of both phenomena separately, before we combine them for understanding the overall pressure evolution.  
The pressure development of the superconducting state is mainly governed by two features, which get directly visible in Figure \ref{figure9} (a). In the high pressure region above $14$~kbar, the superconducting transition becomes incomplete until at $16$~kbar, only a weak kink is observed, without a clear superconducting downturn. We attribute this kink to a change of electronic scattering due to the Eu$^{2+}$ order and not to superconductivity, as we will explain later in the context of the overall phase diagram. 

\begin{figure}
\centering
\includegraphics
[width=0.9\columnwidth]
{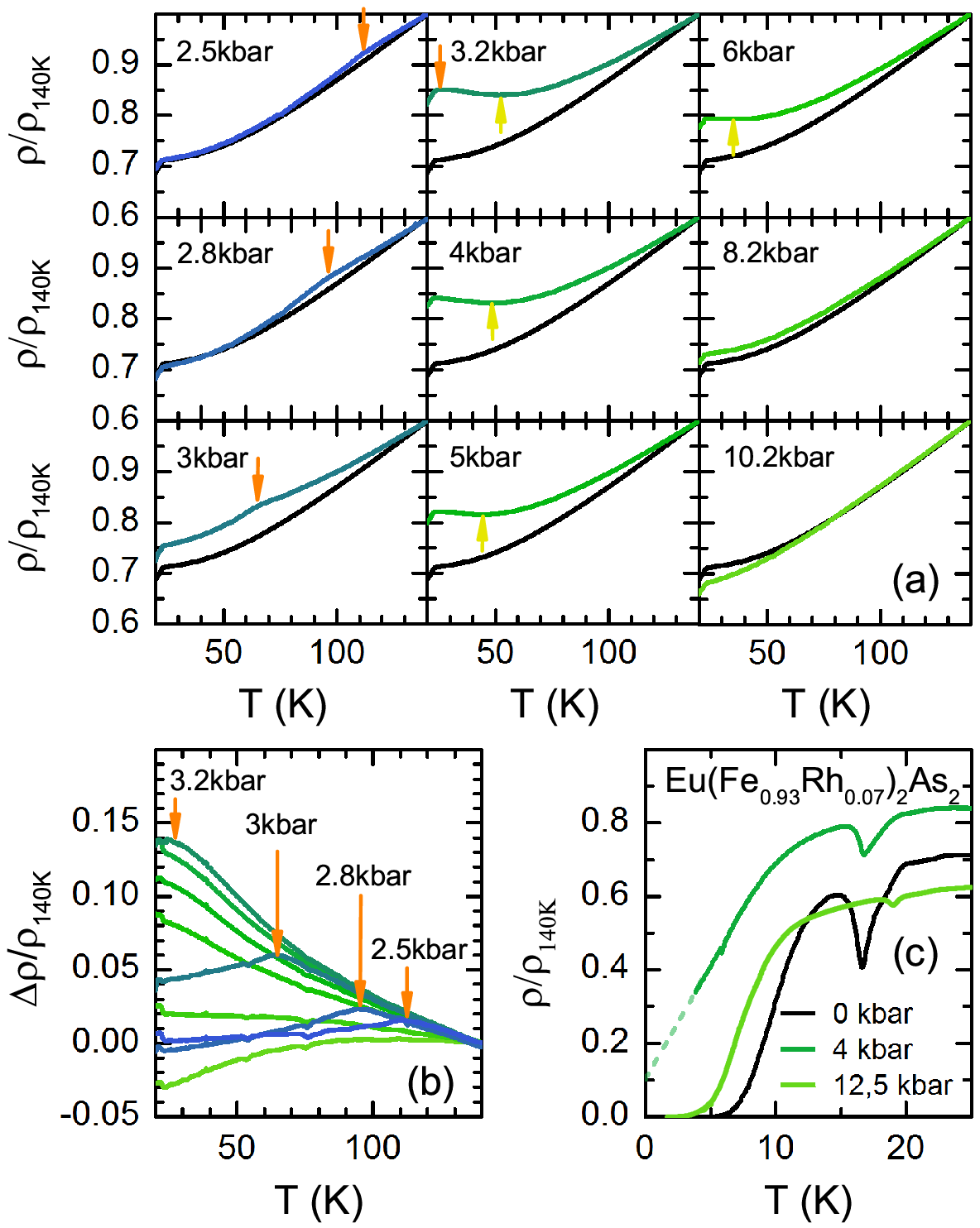}
\caption{(Color online) Pressure dependent resistivity measurements of \Eu \ in the intermediate pressure range. (a) Normalized resistivity plotted in comparison to the ambient pressure resistivity (black curves) to illustrate the appearance of the additional feature above the superconducting phase. (b) Differences of the pressurized resistivity to the ambient pressure data.The orange arrows indicate the temperatures $T_{\rm{F}}$ where the system shows the strongest deviation from the $0$~kbar data. (c) Low temperature data before, during and after the appearance of the resistivity feature; showing a clear broadening in the presence of the feature whereas a sharp transition recovers at higher pressures restoring $T_{c,0}$.} 
\label{figure10}
\end{figure}

In the intermediate pressure region ($2.5$~kbar - $8.2$~kbar), an additional feature becomes visible in the resistivity curve $\rho(T)$, indicated by the orange arrows in Figure \ref{figure9} (a). To follow the evolution of this feature, a comparison of the normalized resistivity to the ambient pressure data is depicted in Figure \ref{figure10}. 
Starting with a small hump at around $110$~K at $2.5$~kbar, it shifts with increasing pressure down to lower temperatures; there it even leads to a strong resistivity upturn, which gets more pronounced the closer it approaches $T_{\rm{c,on}}$.
This behavior has a big impact on $T_{\rm{c,0}}$ and results in a strong suppression of $T_{\rm{c,0}}$ around $5$~kbar.  
In contrast, $T_{\rm{c,on}}$ is very robust against pressure exposure and  stays constant up to $14$~kbar, until the superconductivity gets rapidly suppressed. In order to follow the changes of the magnetic phases with hydrostatic pressure, we plot the shifted resistivity curves of \Eu \ for different values of hydrostatic pressure up to $12.5$~kbar (at room temperature) in Figure \ref{figure9} (c). With increasing pressure, the resistivity dip related to the antiferromagnetic ordering of the Eu$^{2+}$ moments, shifts to higher temperatures, as observed in Ref. \onlinecite{Zapf16} for isovalent substitution and electron doping. 
By analyzing the data with considering the pressure loss during the cooling process, we find a linear increase with a rate of $dT_{\rm{N}}/dp =0.22\pm0.01$~K/kbar. On the basis of the ambient pressure analysis above, we expect the SG transition to be located close to the local maximum of $\rho(T)$. Therefore, we conclude that the spin glass transition (at $T_{\rm{max}}$) also increases with pressure, however slightly faster than $dT_{\rm{N}}/dp$.

\subsubsection{Normal state properties}
For the analysis of the normal state properties, the power law fitting from Equation (\ref{eq:rho}) was employed for temperatures above $T_{\rm{c,on}}$, or alternatively above $T_{\rm{N}}$ after the suppression of the superconducting state. The evolution of the exponent $n$ with pressure is plotted in Figure \ref{figure9} (b). Here one can see that $\rho(T)$ changes from a quadratic temperature dependency at ambient pressure to a more linear one with increasing pressure. At $p=18$~kbar, the resistivity indeed shows a linear progression in the whole measured temperature region above $T_{\rm{N}}$, as can be seen also directly by the black dashed line in Figure \ref{figure9}. This indicates that we have a crossover from a Fermi-liquid, corresponding to an exponent $n=2$,  at low pressures to a non-Fermi-liquid like regime ($n=1$) at high pressures. Surprisingly, it occurs at the boundary of the superconducting phase and not at maximum $T_{c}$, as observed typically in 122 iron pnictides \cite{Shibauchi14, Abrahams11}. One possible explanation for the pressure induced non-Fermi-liquid behavior at the edge of the superconducting dome is given in Ref. \onlinecite{Dai15}, where they observe a similar evolution of the temperature dependency of $\rho$ by Co doping in LiFeAs.
Here, low-energy spin fluctuations tuned by Fermi-surface nesting could be identified as the cause of the crossover to a non-Fermi-liquid regime. To pin down this scenario in \Eu, additional measurements such as nuclear magnetic resonance (NMR), are needed to identify the low-energy spin fluctuation strength under pressure.\\
A second scenario is the following: in our compound, the linear temperature dependence of $\rho$ originates from a quantum critical point. However, the high Eu ordering temperature (which increases with pressure) suppresses $T_c$. Therefore, the quantum critical point is somehow masked by the Eu ordering.

\subsubsection{Pressure-dependent phase diagram of \Eu}
\begin{figure}
\centering
\includegraphics
[width=0.9\columnwidth]
{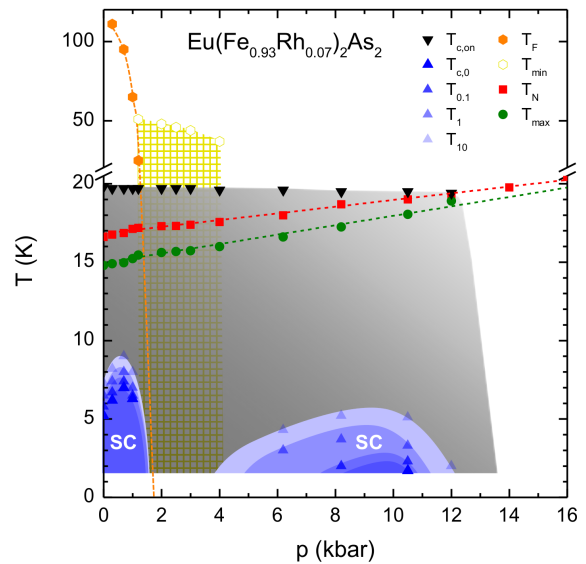}
\caption{(Color online) Pressure-dependent phase diagram of \Eu \ generated from the resistivity measurements depicted in Figure \ref{figure9} (pressures are corrected to the pressure loss during the cooling process). The magnetic transition $T_{\rm{N}}$ (red squares) increases linearly with increasing pressure, while the superconducting phase (black triangles) gets completely suppressed with pressures around $14$~kbar. The local maximum of the re-entrant transition, which is probably caused by the SG state, is depicted in green and shows a bit faster increase with pressure than $T_{\rm{N}}$. The zero resistance point $T_{\rm{c,0}}$ as well as $T_{0,1}$, $T_{1}$ and $T_{10}$ (representing the temperatures were the resistivity drops below $0.1$~$\mu\Omega$cm, $1$~$\mu\Omega$cm and $10$~$\mu\Omega$cm respectively) are labeled with blue triangles, exhibit an anomaly in the intermediate pressure range connected to the appearance of the feature at higher temperatures $T_{\rm{F}}$ (orange hexagons). Additionally the temperature range were a local minimum of  $\rho(T)$ appears above the superconducting transition is marked in yellow.}  
\label{figure11}
\end{figure}

In Figure \ref{figure11} all pressure dependent measurements are summarized in a pressure-temperature-phase-diagram. The labeled pressure values are thereby corrected with regard to the pressure loss during the cooling process.

In this diagram, two processes, which cause the suppression of superconductivity, become very obvious: First of all, any trace of superconductivity is suppressed at the point where the Eu ordering temperatures get in close proximity to $T_{\rm{c,on}}$. This demonstrates clearly the strong competition of superconductivity and magnetism in \Eu. 
Secondly, the high-temperature anomaly observed at intermediate pressures is strongly interacting with $T_{\rm{c,0}}$. The direct influence of the feature to the superconducting phase is not so straight forward as in the first case, instead a multifaceted picture opens up.
At intermediate pressures the kink-like feature at $T_{\rm{F}}$ shifts down rapidly with pressure while $T_{\rm{c,0}}$ slightly increases up to $7$~K at $0.7$~kbar. In the pressure range from $2$~kbar to $4$~kbar (yellow shaded area), where the high temperature feature indicates its continuous presence in an upturn of $\rho(T)$ [Fig. \ref{figure10} (a), yellow arrows], the superconducting transition gets significantly broader [Fig. \ref{figure10} (c)], keeping the system from reaching zero resistance. Only after the influence of the high temperature feature is completely suppressed at pressures around $4$~kbar, the full superconducting transition recovers and a second dome arises which exists till superconductivity gets completely suppressed. 
This coupled development of $T_{\rm{F}}$ and $T_{\rm{c,0}}$  indicates a strong competition of the two phases. In order to identify its origin, additional measurements under applied external magnetic fields were employed which will be discussed in the following paragraph.

\subsubsection{Magnetic field dependences of pressurized \Eu}
\begin{figure}
\centering
\includegraphics
[width=0.9\columnwidth]
{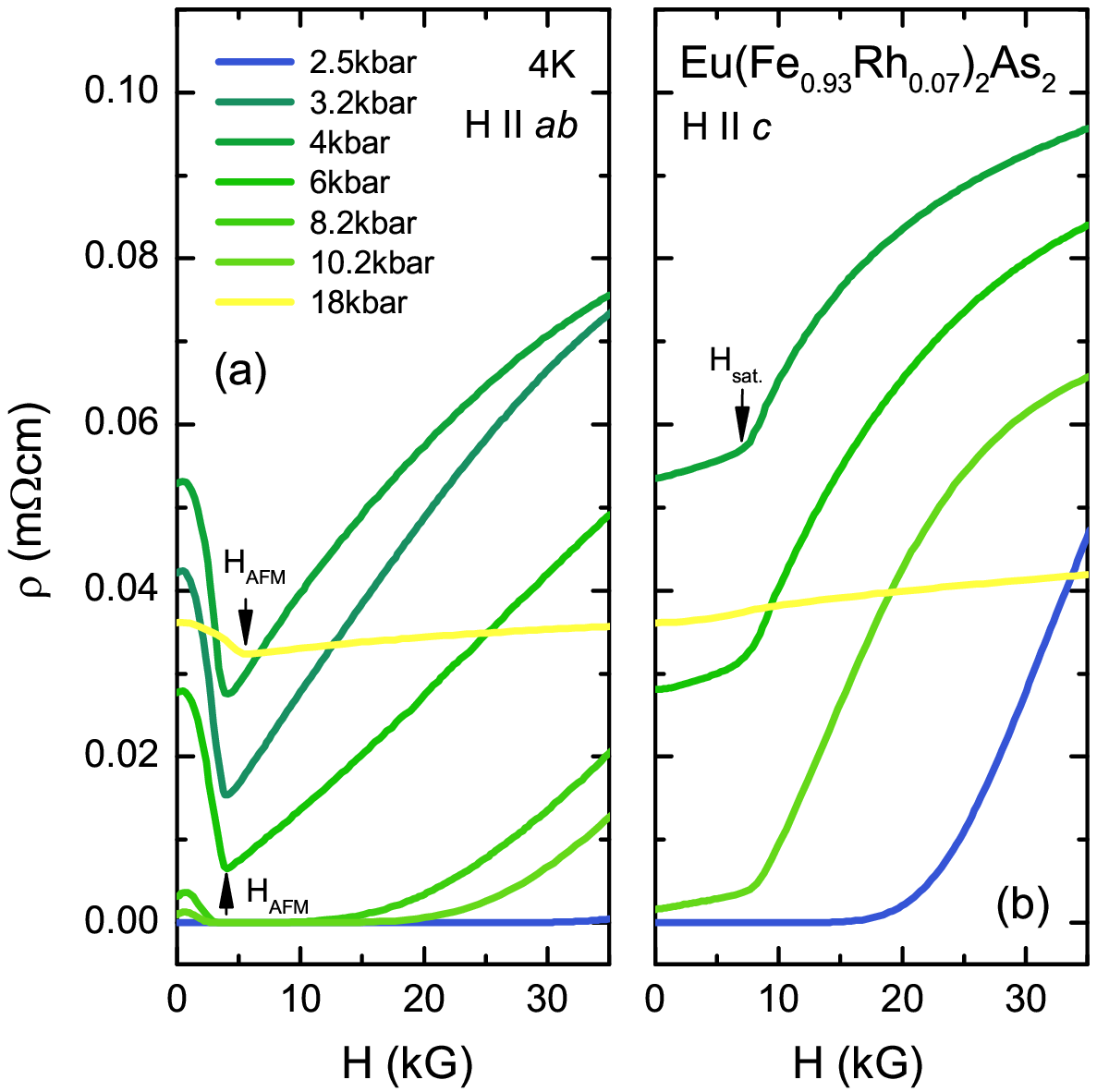}
\caption{(Color online) Isothermal resistivity measurements under magnetic fields applied perpendicular (left panel) and parallel (right panel) to the \textit{ab}-plane of \Eu \ at different hydrostatic pressures up to $18$~kbar (room temperature pressure values). While for the out-of-plane configuration a purely positive slope can be observed, the in-plane measurements exhibit a negative slope at low fields, connected with the field induced enhancement of $T_{\rm{c,0}}$. Two critical fields manifest themselves in the data (a) as a minimum in the in-plane configuration at $H_{\rm{AFM}} \approx 4$~kG and (b) as a kink in the out-of-plane measurement at $H_{\rm{sat}} \approx 8$~kG (see the black arrows).}
\label{figure12}
\end{figure}

Figure \ref{figure12} depicts isothermal in- and out-of plane magnetoresistance measurements at $4$~K for different hydrostatic pressures up to $18$~kbar (room temperature pressure values). A clear anisotropic behavior becomes visible, as the out-of-plane field induces a purely positive slope of the resistivity curve, while the in-plane magnetic field first reduces the resistance considerably up to a critical field $H_{\rm{AFM}}$ (right panel, black arrows), where the slope changes its sign and finally a positive magnetoresistance settles at high fields. Also in the out-of-plane data, a critical field $H_{\rm{sat}}$ (left panel, black arrow) can be identified as a clear kink with $H_{\rm{sat}} >H_{\rm{AFM}}$.

\begin{figure}
\centering
\includegraphics
[width=0.9\columnwidth]
{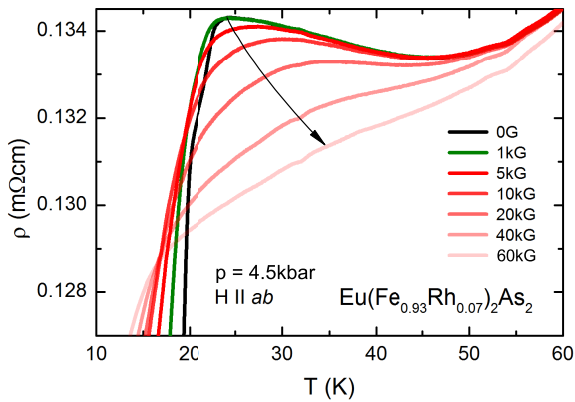}
\caption{(Color online) Temperature dependent magnetoresistance measurements at $4.5$~kbar show the evolution of the pressure induced cusp in $\rho(T)$ at around $25$~K with in-plane magnetic fields. The local maximum gets broadened and shifts to higher temperatures till at fields of $40$~kG the upturn is no longer present in the resistivity curve.}
\label{figure13}
\end{figure}

The underlying physical processes that lead to the big changes in resistivity at the critical fields can be identified by comparing the critical fields with the temperature dependent magnetoresistance measurements at ambient pressure, as well as the magnetization measurements.
The in plane-critical field of around $H_{\rm{AFM}} \approx 4$~kG at $3.2$~kbar coincides with the suppression of the antiferromagnetic ordering temperature, as can be seen best in the upper phase diagram of Figure \ref{figure8}. Moreover, the shift of $H_{\rm{AFM}}$ to slightly higher fields with pressure ($H_{\rm{AFM}} \approx 5$kG at $18$~kbar) reflects the enhancement of $T_{\rm{N}}$ by pressure.
Thus this behavior shows clearly the relationship of negative magnetoresistance and the enhancement of $T_{\rm{c,0}}$ with the magnetic ordering in \Eu. 
For the out-of-plane measurements, the critical field $H_{\rm{sat}} \approx 8$~kG corresponds to the onset of the magnetic saturation, as can be concluded from the magnetization data of Figure \ref{figure7}. 
The large positive magnetoresistance at intermediate fields - after the suppression or saturation of all internal magnetic effects - occurs in both cases due to paramagnetic and orbital pair-breaking effects induced by the external magnetic field, as expected for any normal superconductor without magnetic ordering phenomena. At high fields, a saturation of the magnetoresistance sets in as soon as the superconducting phase is completely destroyed. \\
To get an idea about the origin of the up to now unidentified resistivity feature evolving at low pressures, magnetoresistance measurements were carried out in a temperature range from $4$~K up to $60$~K at a pressure of $4.5$~kbar, where the feature is especially pronounced. Figure \ref{figure13} shows the resulting resistivity curves obtained with applied magnetic fields in-plane (out-of-plane magnetic fields induce the same field dependence). As indicated by the black arrow, magnetic fields lead to a broadening and suppression of the resistivity hump. Therefore, this feature is probably of magnetic nature. \\
In principle, there are two possible scenarios to explain the origin of this feature: the Kondo effect and the spin density wave. 
As the feature appears at low pressures already at high temperatures and shows a very similar curvature compared to the typical SDW in pressurized EuFe$_2$As$_2$ \cite{Kurita13} and doped BaFe$_2$As$_2$ \cite{Rullier-Albenque16}, we suggest that external pressure re-activates the SDW in \Eu. This view is supported by the phase diagram (Fig. \ref{figure11}) revealing a suppression of $T_{\rm{c,0}}$ in the same pressure range, indicating the well known competition of the two phases in the iron pnictide family.
With regard to the general phase diagram of the EuFe$_2$As$_2$ parent compound, the re-activation can be caused by a competition between the effects of the applied external pressure and the electron doping which comes, in case of Rh, along with a negative internal pressure effect \cite{Jeevan08, Singh09}. 
By taken into account that not just doping and the unit cell volume are parameters to tune the high-$T_c$ superconductivity and magnetic phases in layered iron pnictides, but also peculiarities of the crystal structure, like the tetrahedral angel play a crucial role \cite{Johnston10}, the combination of internal and external pressure can lead to configurations not accessible by just one tuning-parameter.

As the compound seems to be close to Fermi-surface nesting instabilities, small external pressure can counteract to the internal negative pre-pressure paving the way for the appearance of a SDW in the under-doped regime.
To pin down this scenario, further investigations are needed, preferably magnetic susceptibility and neutron scattering measurements under pressure would be a powerful tool. 
Additionally further pressure dependent measurements on chemically substituted superconducting compounds with negative pre-pressure, such as Eu(Fe$_{1-x}$Ir$_x$)$_2$As$_2$ or other members of the Rh doped family, would be interesting to see, if significantly higher $T_c$ values can be reached in this way.

\section{Conclusion and Outlook}
\label{conclusion}
High-pressure magneto-transport measurements up to $18$~kbar have been performed on \Eu \ single crystals to establish a pressure-dependent phase diagram of an electron doped re-entrant superconducting member of the EuFe$_2$As$_2$ iron-pnictide family. The systematic variation of the two parameters pressure and magnetic field provides thereby the possibility to vary individual aspects of the complex system to get an insight to the interesting interplay of superconductivity and magnetism.
The achieved magnetic field depended phase diagram at ambient pressure reveals the validity of the picture of a canted A-type antiferromagnetic ordering which even leads to a field-induced enhancement of $T_{\rm{c,0}}$ in the order of $2$~K. Magnetization measurements additionally confirm the existence of a spin-glass phase and uncover the coexistence of ferromagnetism and superconductivity at low temperatures. Thus, \Eu \ can be count to the family of  ``magnetic superconductors''. 
The thermal hysteresis in $\rho(T)$ around the superconducting transition was identified as a novel helpful tool to follow the magnetic and superconducting transition temperatures under applied magnetic fields.
The pressure dependent measurements show a fast suppression of the superconducting state at a pressure around $14$~kbar where $T_{\rm{N}} \approx T_{\rm{c,on}}$, 
opening the view on the direct competition between the two phenomena.  
By contrast to other iron-based superconductors -- showing non-Fermi liquid behavior above the superconducting dome connected with a hidden quantum critical point -- we observe a $T$-linear development of the resistivity at the edge of the superconducting dome. 
Furthermore, an additional magnetic field dependent feature develops in the intermediate pressure range.  We identify it as a SDW reactivated by pressure, which counteracts the negative internal pressure induced by chemical substitution. This indicates once more that not only doping influences the phase diagram; instead, small variations of the lattice by internal and external pressure plays an important role. Moreover, we open with our experiments a novel path to manipulate and eventually enhance $T_c$ in compounds under negative internal pressure by additional external pressure.

\acknowledgments
We thank C. Kamella for performing the EXD measurements at the MPI in Stuttgart, G. Untereiner for preparing the crystals and D. V. Efremov for fruitful discussions . We acknowledge support by the Deutsche Forschungsgemeinschaft (DFG) SPP 1458.

\end{document}